\documentclass{aastex}

\slugcomment{submitted to The Astrophysical Journal, 30/jan/2000}

\shorttitle{Masi et al.}
\shortauthors{Galactic dust in the BOOMERanG maps}

\def\ltsima{$\; \buildrel < \over \sim \;$}
\def\simlt{\lower.5ex\hbox{\ltsima}}
\def\gtsima{$\; \buildrel > \over \sim \;$}
\def\simgt{\lower.5ex\hbox{\gtsima}}

\begin{document}

\title{High latitude Galactic dust emission in the BOOMERanG maps}

\author{S. Masi\altaffilmark{1}, P.A.R. Ade\altaffilmark{2},J.J Bock\altaffilmark{3,4}, A. Boscaleri\altaffilmark{5}, B.P. Crill\altaffilmark{3}, 
P. de Bernardis\altaffilmark{1}, M. Giacometti\altaffilmark{1}, 
E. Hivon\altaffilmark{3}, V.V. Hristov\altaffilmark{3}, 
A.E. Lange\altaffilmark{3}, P.D. Mauskopf\altaffilmark{6}, T. Montroy\altaffilmark{7}, 
C.B. Netterfield\altaffilmark{8}, E. Pascale\altaffilmark{5}, F. Piacentini\altaffilmark{1}, 
S. Prunet\altaffilmark{8}, J. Ruhl\altaffilmark{7} }

\affil{\altaffilmark{1} Dipartimento di Fisica, Universita' La Sapienza, Roma, Italy}

\affil{\altaffilmark{2} QMWC London UK}

\affil{\altaffilmark{3} CALTECH, Pasadena, USA}

\affil{\altaffilmark{4} JPL, Pasadena, USA}

\affil{\altaffilmark{5} IROE-CNR, Firenze, Italy}

\affil{\altaffilmark{6} Department of Astronomy, Univ. of Wales, UK}

\affil{\altaffilmark{7} Department of Astronomy, Univ. of California at Santa Barbara, USA}

\affil{\altaffilmark{8} Department of Astronomy, Univ. of Toronto, Canada}

\begin{abstract}

We present mm-wave observations obtained by the BOOMERanG
experiment of Galactic emission at intermediate and high 
($b < -20^o$) Galactic latitudes. We find that this emission is 
well correlated with extrapolation of the IRAS-DIRBE maps, and is spectrally 
consistent with thermal emission from interstellar dust (ISD).
The ISD brightness in the 410 GHz map has an angular power spectrum 
$c_\ell \sim \ell^{-\beta}$ with $2 \simlt \beta \simlt 3$. 
At 150 GHz and at multipoles $\ell \sim 200 $ 
the angular power spectrum of the IRAS-correlated dust signal is estimated to be
$\ell (\ell + 1) c_\ell / 2 \pi = (3.7 \pm 2.9) \mu K^2$. 
This is negligible with respect to the CMB signal
measured by the same experiment
$\ell (\ell + 1) c_\ell / 2 \pi = (4700 \pm 540) \mu K^2$.
For the uncorrelated dust signal we set an upper limit to the
contribution to the CMB power at 150GHz and $\ell \sim 200 $ of
$\ell (\ell + 1) c_\ell / 2 \pi < 3 \mu K^2$ at 95$\%$ C.L. .

\end{abstract}


\section{Introduction}

The patchy emission of our Galaxy 
is a major concern for experiments designed to measure the
anisotropy of the Cosmic Microwave Background (CMB).
A "precision" phase, where temperature 
fluctuations are measured with a sensitivity of the order
of tens of $\mu K$ per pixel, has now begun \cite{debe2000, Hana2000}. 
Forthcoming full-sky coverage 
space missions \cite{map, planck}, and a host
of future sub-orbital experiments are expected to reach 
sensitivities of a few $\mu K$ per pixel .  
To fully exploit the potential of these new surveys, 
our knowledge of the diffuse emission of our Galaxy at high 
Galactic latitudes must improve as well. 

At frequencies above
$\sim$ 100 GHz this emission is dominated by 
thermal radiation from large dust grains, heated by the interstellar
radiation field to $T_d \sim 10 - 30 K$.
The ISD is distributed in filamentary "cirrus"-like clouds and 
covers the sky even at high Galactic latitudes \cite{Low84}.
The spectrum of this component in the 300 - 3000 GHz range has 
been mapped with coarse (7$^o$) angular resolution and high sensitivity by the 
COBE-FIRAS experiment \cite{Wrig91}. The COBE-DIRBE maps provide higher angular
resolution $\sim 0.7^o$,
at $>$ 1250 GHz \cite{Haus98,Aren98}. Arcminutes 
resolution maps from the IRAS satellite are available only at $>$ 3000 GHz.
These have been recalibrated using the 
COBE-DIRBE maps at 3000 and 1250 GHz \cite{Schl99},
and extrapolated to longer wavelengths using a variety of
physical models \cite{Laga98, Fink99, Laga00, Tegm00}.
At those longer wavelengths very few 
experimental data are available at subdegree resolution 
\cite{Masi95, Masi96, Lim96, Leit97, deOl97, Che97}.
"Anomalous" emission, morphologically correlated with the
IRAS map but much larger than a naive 
extrapolation of thermal dust emission, has been detected 
in the microwaves \cite{Kogu96a, Kogu96b, Lim96, Leit97, 
deOl97, deOl98, Mukh99, Drai98, deOl00}.

Here we analyze 90, 150, 240 and 410 GHz maps of 3$\%$ of the sky at Galactic 
latitudes $10^o \simlt b \simlt 60^o$. 
We compute the frequency and angular power spectrum 
of the fluctuations in these maps. We find that these maps are correlated with 
the emission mapped by IRAS extrapolated to our wavelengths using \cite{Fink99}
 model number 8. We will refer to this as FDS8 in the following.
We also set upper limits to the level of residual, non-CMB structures, that are 
not correlated with FDS8.

\section{Observations}

We use the maps obtained from the 1998 long duration flight of the
BOOMERanG experiment \cite{debe2000},
all pixelized with 7' pixels and smoothed to a resolution of 22.5 arcmin FWHM.
The instrument was calibrated against the CMB dipole at 90, 150, 240 GHz (10$\%$
uncertainty), and against the rms CMB anisotropy at 410GHz ($20\%$ uncertainty) 
\cite{Cril2000}. The conversion factors from CMB temperature fluctuations 
to brightness in our 4 bands are $195, 426, 471, 204 (MJy/sr)/K_{CMB}$ at
90, 150, 240 and 410 GHz respectively.
We used a single channel at 90 and at 410 GHz, and combined
3 channels at 150 GHz and 3 channels at 240 GHz. 
About 1300 square degrees were observed 
at high galactic latitudes ($-60^o \simlt b \simlt -20^o$;  
 $230^o \simlt \ell \simlt 270^o$; 
constellations of Caelum, Doradus, Pictor, 
Columba, Puppis), including a region with the lowest amount 
of dust emission of the full sky. 
In the observed region the fluctuation of the 100 $\mu$m brightness 
mapped by IRAS is well below 1 MJy/sr in over 500 square degrees.
The BOOMERanG maps have been obtained from the raw data using
an iterative algorithm \cite{Prun2000}. This reduces the large 
scale artifacts due to 1/f noise, and correctly estimates 
the noise in the datastream, while producing a 
maximum likelihood map. 
Structures at scales larger than 10$^o$ are 
effectively removed in the process. This fact must be taken into 
account when comparing the BOOMERanG maps to other maps of the sky.
In addition to the four frequencies mapped by BOOMERanG, we use 
the FDS8 dust maps as explained in next section. 

\section{The 410 GHz "dust monitor"}

The highest frequency channel of the BOOMERanG photometer is centered
at 410 GHz, with a FWHM of 26 GHz. At this frequency, the brightness of the
CMB is smaller than at our lower frequencies, while thermal emission from 
Galactic dust is much larger. The 410 GHz map of the sky obtained by
BOOMERanG is dominated by faint cirrus clouds at intermediate Galactic
latitudes ($-10^o < b < -20^o$). In fig.1 we compare our 410 GHz map (top panel) to 
the FDS8 extrapolation of the IRAS map (middle panel) obtained as follows.  
FDS8 assumes two components
of ISD with different temperature and spectral index of
dust emissivity. The two temperatures depend on the observed direction.
On the average $<T_{d,1}> \sim 16.2 K, <T_{d,2}> \sim 9.4 K$, and 
the average ratio between 
dust brightness at 3000 GHz and dust 
brightness at 410 GHz is $\sim 13$.
The extrapolated map has been sampled along the scans of the
410 GHz channel of BOOMERanG, and then high-pass and low-pass 
filtered using the 410 GHz detector transfer function, 
in order to create a synthesized time-stream.
The time stream has been processed in the same way as the 
BOOMERanG data, and smoothed to 22.5 arcmin to obtain the map shown in 
the middle panel of fig.1. 
The morphological and amplitude agreement of the two maps 
provides evidence that the 410 GHz data represent a good
 monitor for interstellar dust emission in the BOOMERanG data.
Bright compact structures apparent in the difference map
(lower panel in fig.1) correspond to the dense, cool cores of clouds
that are not well modelled in FDS8. The remaining structures 
are mostly due to residual noise in the BOOMERanG data.
In a similar way we obtained
extrapolated maps for the other BOOMERanG channels.

We have computed the power spectrum of the 410 GHz map in
three circular regions, each $18^o$ in diameter, centered at
(RA, dec, b) = (107$^o$, -47$^o$, -17$^o$), (92$^o$, -48$^o$, -27$^o$), 
(74$^o$, -46$^o$, -38$^o$) i.e. low, intermediate and high Galactic 
latitudes respectively. We used a spherical harmonics transform and
corrected for the finite size of the cap, for filtering applied
in the time domain, and for the contribution of instrumental noise \cite{Hiv01}. 
The results are shown in figure 2.
The contribution of CMB anisotropy to these spectra is computed to
be negligible. The errors have been computed by adding two contributions. 
The first one is an estimate of instrumental noise. 
The second is an estimate of sampling variance \cite{Sco93}
for a gaussian field having the same power spectrum. The latter 
has to be included if we want to consider the measured spectrum
as representative of ISD fluctuations in the Galaxy in general.
The spectrum at the highest Galactic latitude is basically 
an upper limit for dust fluctuations,
since the residual fluctuations are comparable to our estimate
of detector noise plus CMB anisotropy.
The spectra at low and intermediate latitudes 
are well fit by a power law $c_l \sim \ell^{-\beta}$
as in previous studies based on IRAS and DIRBE maps
\cite{Gaut92, Low94, Guar95, Wrig98, Schl99}.
We find a power law exponent $2 \simlt \beta \simlt 3$, consistent with
the studies cited above, thus extending this result to
wavelengths very close to those used for CMB studies.
The power spectra of FDS8 at 410 GHz in
the same regions are also shown in fig.2 for comparison. The agreement
is very good for the region centered at $b = -17^o$, where
detector noise is negligible. The agreement is also good in the region
centered at $b = -27^o$, but a systematic amplitude difference is evident.
We estimate upper limits
for the fluctuations due to any dust component not correlated
with IRAS by computing the spectrum of the difference map obtained 
removing the F8 map from the measured 400GHz map. The upper limits are of the
same order of magnitude of the errors in the measured power
spectrum of the 400 GHz map. 

\section{Pixel-pixel correlations}

We made pixel-pixel correlations between our four maps 
and the corresponding FDS8 maps. The signal in 
each of our channels is a linear combination of Galactic emission, 
CMB anisotropies and noise. The relative weight of the Galactic and the CMB
components depends on the Galactic latitude and on the frequency of the
channel. The advantage of correlating with the FDS8 maps is that the noises
are uncorrelated, and at 3000 GHz the CMB is totally negligible. 
Any detected correlation is thus due to Galactic emission.
In the BOOMERanG 410 GHz channel we expect to have little 
CMB anisotropy and dominant Galactic dust emission, at least
at $b > -20^o$. In fact, in this latitude range the
pixel-pixel scatter plot of our 410 GHz channel vs FDS8 at 410 GHz has a 
best fit line with slope $(0.644 \pm 0.038)$, 
a highly significant correlation. 
This result has been obtained using a jack-knife 
technique: we divide the latitude band $-20^o < b < -10^o$ into five
$10^o \times 10^o$ regions and we compute the best fit slope 
for each of the regions. We then compute 
the average and standard error on the average as our 
best estimate of the general slope. 
In this way we properly take into account the fact
that deviations from an ideal correlation are dominated by fluctuations
in dust properties, rather than by detector noise.
The slopes and Pearson's linear correlation coefficients 
are listed in table 1.
As we move towards lower frequencies, the correlation at a given
latitude range gets worse, but is still significant.
We have converted the measured slopes into brightness ratios
$R_i = \Delta B_i / \Delta B_{IRAS}$
using the spectral response of the BOOMERanG bands.
The spectrum of the brightness ratios is plotted in fig.3 (triangles).
We compare it to an empirical model assuming a power law
$B(\nu) \sim \nu^{\alpha}$. We find a best fit $\alpha = (3.2 \pm 0.3)$ at $b > -20^o$. 
In FDS8 model, $\alpha = 3.15$ in the range 240-410GHz, while
$\alpha = 3.36$ in the range 240-150GHz. 

At higher Galactic latitudes (four $10^o$ wide 
latitude bands at $b < -20^o$) we find 
poorer but still significant correlations 
at 410 GHz and 240 GHz, while the correlation is just 
marginal at 150 GHz, and is negligible at 90 GHz (see table 1 and squares
in fig.3). Here we get a best fit $\alpha = (4.3 \pm 1.0)$.

%Dust emission can be naively modeled as
%a Planck spectrum times a power law emissivity:
%$B(\nu) \sim \nu^{\gamma} BB(\nu, T_d)$. The 
%"canonical" range for $\gamma$ is $1 \simlt \gamma \simlt 2$ 
%\cite{Poll94}. The value of $\alpha$ we measure is 

\section{Contamination of CMB measurements at high Galactic latitudes}

We can use the measured ratios
$R_i$ to estimate the rms fluctuation due to IRAS-correlated
emission at high Galactic latitudes. We have simply
$var(\Delta B_i) =  R_i^2 var(\Delta B_{IRAS})$.
%These are conservative estimates of the contamination
%expecially at high Galactic latitudes, since 
%$var(\Delta B_{IRAS})$ contains a significant
%fraction of noise. 
We divide the observed region in latitude 
ranges, 10$^o$ wide, and list the computed mean square fluctuations 
in table 2.
The mean square fluctuation is dominated by signals at the
lowest multipoles, due to the falling power spectrum of dust
$c_\ell \sim \ell^{-2.5}$. So at multipoles corresponding to
the first acoustic peak of the CMB anisotropy the dust
contamination is even more negligible with respect to the CMB 
fluctuations.
We compute the power spectrum in band $i$ with the
simple scaling formula 
$c_{\ell,i} = {R_i^2 \over R_{410}^2} c_{\ell,410} $
where $R_i$ is the average ratio
between the dust signal in band $i$ and the IRAS/DIRBE signal. 
The result is plotted in fig.4 for the extrapolation of 
the 410 GHz spectrum centered
at $b=-27^o$.
Due to the poor correlation, only upper limits are
found for the 90 GHz band (at a level similar to the power
spectrum estimates for the 150 GHz channel), which are not plotted. 
It is evident that at $\ell > 100$
the dust signals at 90 and 150 GHz are negligible with 
respect to the cosmological signal. These estimates of contamination
are consistent with the dust foreground model of \cite{Tegm00} 
(compare their fig.3 with our fig.4). If we scale to 150 GHz the
upper limits for the uncorrelated component, assuming the same
spectral ratios, we get 95$\%$ C.L. upper limits 
$\ell (\ell + 1) c_\ell / 2 \pi \simlt 5 \mu K^2$ at 95$\%$ C.L. 
for $50 \simlt \ell \simlt 600$.

\section{Conclusions}

BOOMERanG has detected thermal emission from insterstellar cirrus at intermediate
and high Galactic latitudes. The 410 GHz map is morphologically
very similar to extrapolation of the IRAS(3000GHz) and DIRBE(1250GHz) maps. 
The angular power spectrum
of the dust dominated 410 GHz map is a power law 
$c_\ell \sim \ell^{-\beta}$ with $2 \simlt \beta \simlt 3$. We have detected 
a component correlated with the IRAS/DIRBE map in all the 
BOOMERanG bands at $-10^o > b > -20^o$, and in the 150, 240 and 410 GHz bands
at higher Galactic latitudes. This dust contamination is negligible with
respect to the CMB anisotropy at high Galactic latitudes, accounting for
less than 1$\%$ of the total angular power spectrum for multipoles $\ell > 100$
at $\nu < 180 GHz$.

\acknowledgments

The BOOMERanG program has been supported by NASA and NSF in the USA, 
by ASI, PNRA, Univrsity La Sapienza in Italy, by PPARC in UK and by NSERC and
University of Toronto in Canada.

\clearpage

%\figcaption[fig1.ps]{ PLATE Top panel: BOOMERanG map at 410GHz. The resolution of the instrument
\figcaption { PLATE Top panel: BOOMERanG map at 410GHz. The 
map has been obtained by coadding the data 
in 7' healpix pixels \cite{Gors98} and applying a gaussian smoothing to an equivalent
resolution of 22.5 arcmin. At this frequency $1 mK_{CMB}$ = 0.20 MJy/sr.
Due to the map-making technique, structures with angular scales larger
than 10$^o$ have been effectively removed from the map. 
Middle panel: FDS8 extrapolation at 410 GHz of the IRAS map, 
in the same sky region. This map has been high and low-pass
filtered as in the BOOMERanG 410 GHz channel (see text)
for a meaningful comparison. Bottom panel: residuals after subtraction 
of the FDS8 map from the BOOMERanG 410 GHz map. 
The three circles identify the regions where we carried out the power
spectrum analysis.
\label{fig1}}

\figcaption[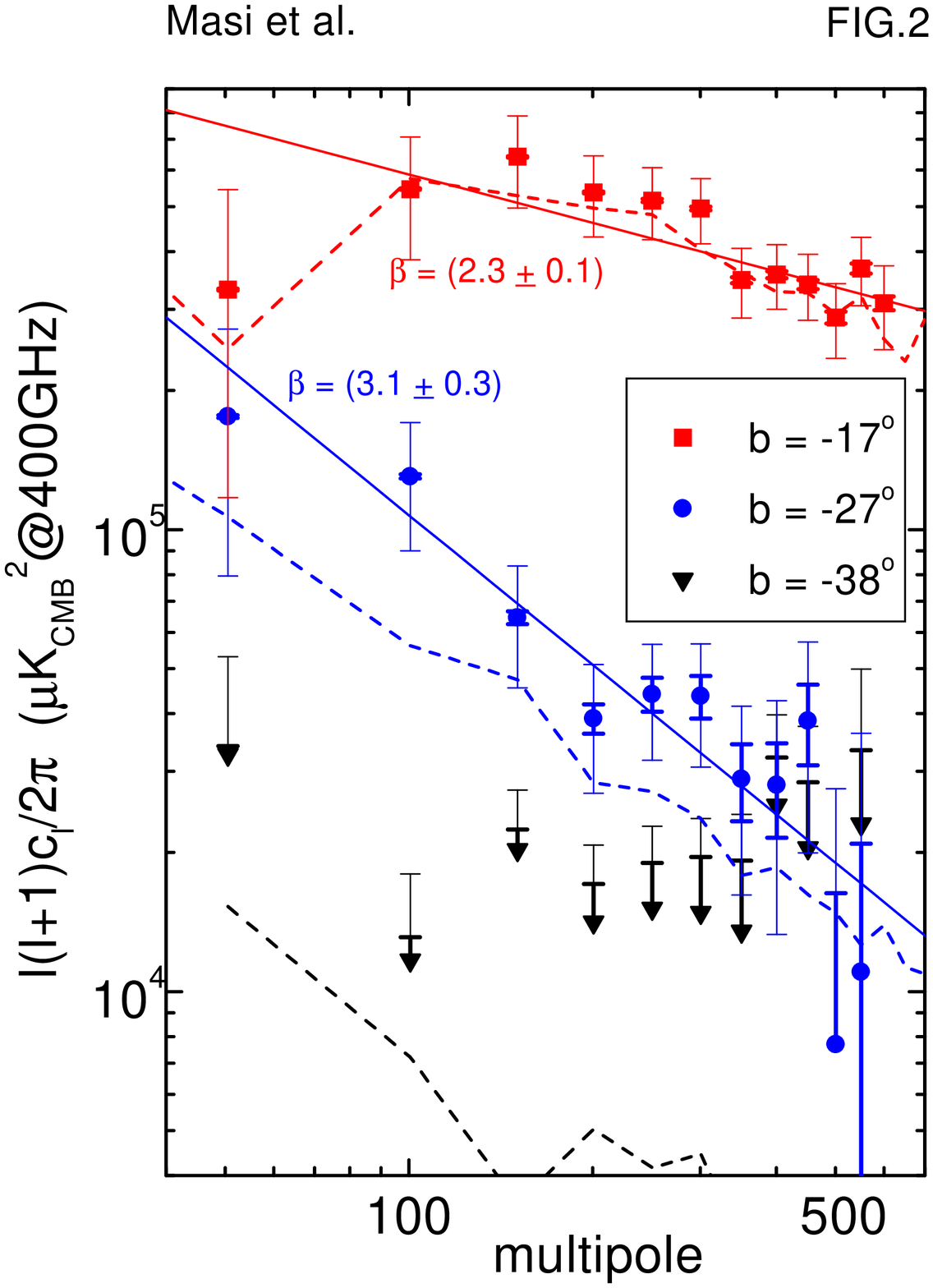]{ Angular power spectrum of the BOOMERanG 410 GHz map, for three
disks with diameter 18$^o$, centered at different Galactic latitudes 
(squares, $b= -17^o$; circles, $b= -27^o$; down triangles, $b = -38^o$).
At the highest latitude the signal is small with respect to the 
detector noise, and we consider the spectrum as an upper limit for
dust brightness fluctuations.
Best fit power-law spectra $c_l \sim \ell^{-\beta}$
are shown as continuous lines and labelled by their best fit slope $\beta$.
The dashed lines are the power spectrum of the FDS8 map at 410 GHz 
in the same sky regions. The large thin error bars include cosmic/sampling
variance, while the smaller thick ones are from intrumental noise only.
\label{fig2}}

\figcaption[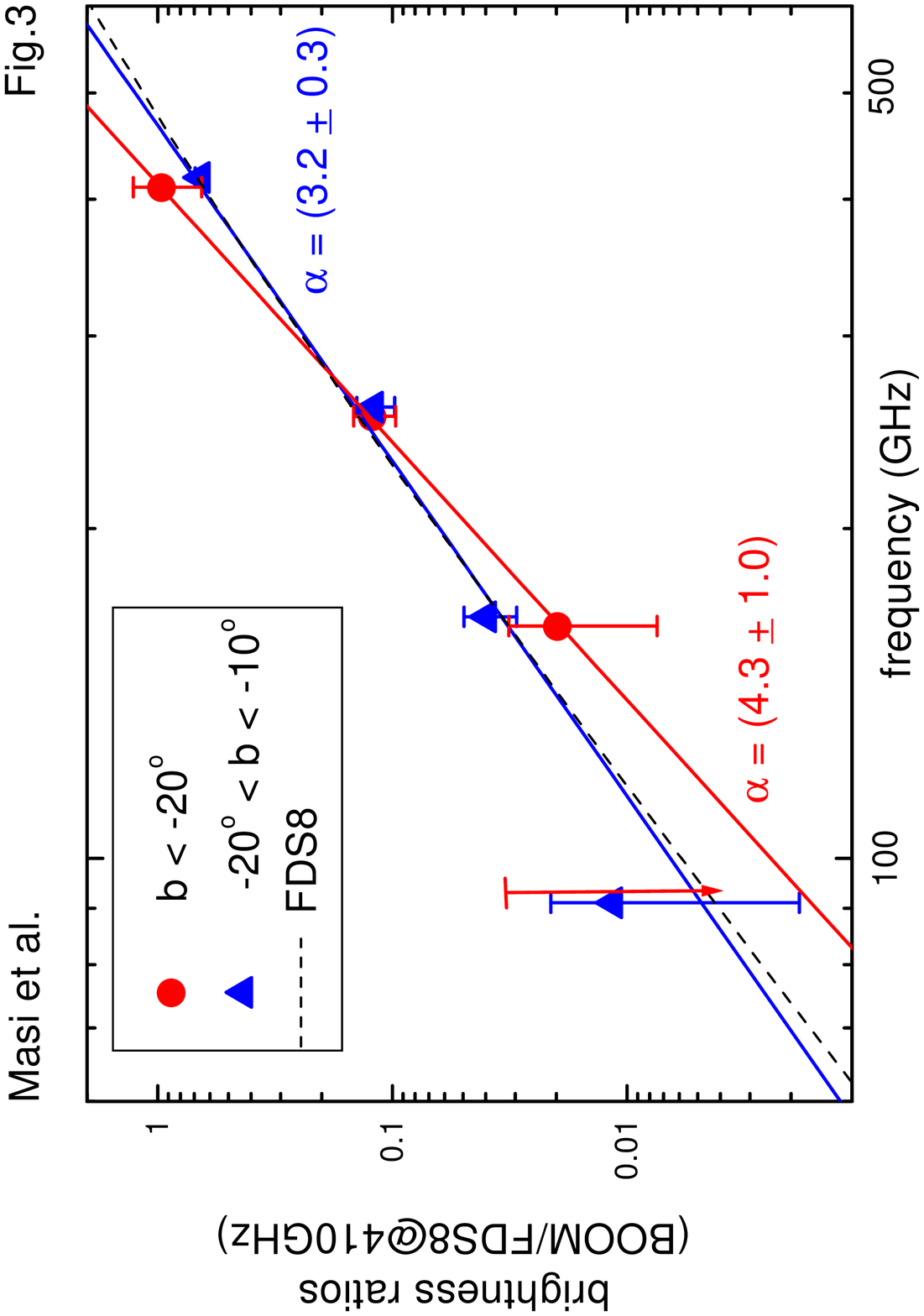]{ Ratio between the dust brightness fluctuations
in the BOOMERanG bands and the dust brightness fluctuations detected 
by IRAS/DIRBE. Triangles indicate measurements at intermediate Galactic
latitudes ($-20^o < b < -10^o$), while circles indicate measurements
at high Galactic latitudes ( $b < -20^o$).
The best fits assuming a power law spectrum with spectral index $\alpha$ 
are plotted as lines, and are labelled by the best fit value for $\alpha$.
The dashed line is the average FDS8 spectrum, normalized to the 
BOOMERanG measurement at 410 GHz at intermediate latitudes.
\label{fig3}}

\figcaption[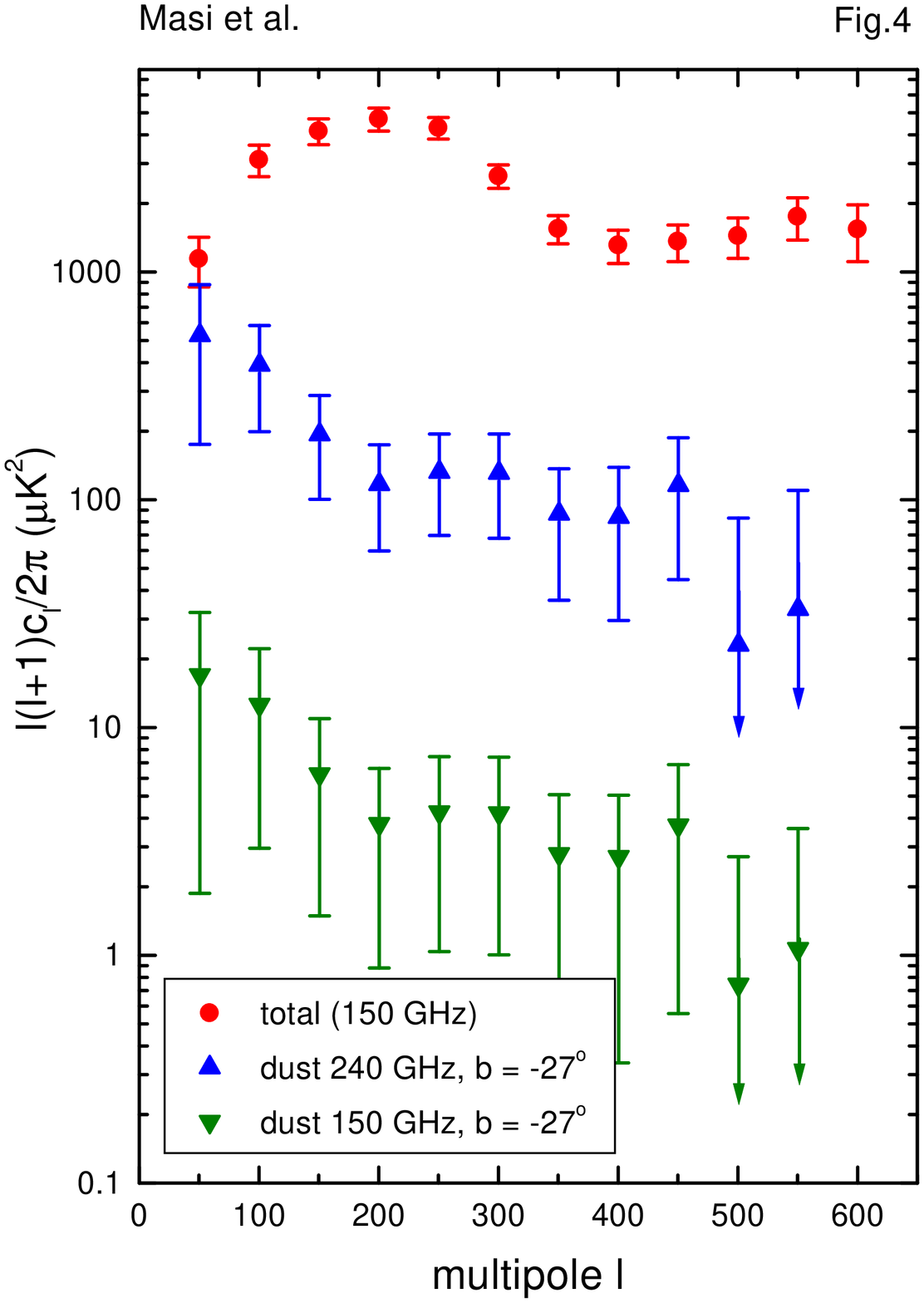]{Angular power spectra of dust emission
in a disk centerd at RA=92$^o$, dec=-48$^o$, b=-27$^o$ that is correlated
with the FDS8 IRAS extrapolation, at 150GHz (filled down triangles) and 
at 240GHz (filled up triangles). The power spectrum of the microwave
sky at 150 GHz \cite{debe2000} is shown for comparison as filled circles.
For the dust spectra the error bars include the uncertainty due to
the partial correlation between our data
and the FDS8 extrapolated IRAS data. The 90 GHz correlation
is so poor that we only find upper limits (not
plotted here) similar to the 150 GHz values.
It is evident that in this region of the sky the Galactic dust signal is
negligible with respect to the CMB signal at these frequencies.
\label{fig4}}

\clearpage

\begin{deluxetable}{crrrr}
%\tabletypesize{\scriptsize}
\tablecaption{Correlation between BOOMERanG data and IRAS/DIRBE data extrapolated to 410 GHz (FDS8). \label{tbl-1}}
\tablewidth{0pt}
\tablehead{
\colhead{Band Center (GHz)} &   
  & \colhead{ slope $R \left[\mu K_{CMB}/(MJy/sr)\right]$ (Pearson's $R$) }  
}
\startdata
     &   $-20^o < b < -10^o$ (22843 pixel) &   $b < -20^o$  (68987 pixel)  \\
     &                             &                           \\
 410 &   $(3200 \pm 190)$ (0.298)  & $(4700 \pm 1500)$ (0.138) \\
%     &   $(0.644 \pm 0.038)$       & $(0.94 \pm 0.30)$         \\
  240 &   $(254 \pm 46)$ (0.156)    & $(258 \pm 52)$ (0.041)    \\
 150 &   $(93 \pm 23)$ (0.085)     & $(46 \pm 29)$ (0.003)     \\
  90 &   $(58 \pm 49)$ (0.032)     & $(-20 \pm 110)$ (-0.028)  \\
\enddata

\end{deluxetable}

\clearpage

\begin{deluxetable}{crrrrr}
\tabletypesize{\scriptsize}
\tablecaption{ 
Estimated rms fluctuations due to dust emission 
$ <\Delta T^2_{dust}>^{1/2}  (\mu K_{CMB})$ (angular scales
between 22 arcmin and  10$^o$).
%Only the fluctuations at scales between 10$^o$ and 22 arcmin 
%(i.e. multipoles $20 \simlt \ell \simlt 600$)
%contribute to the computed rms. Fluctuations outside this
%range have been removed by filters in the time domain and by
%smoothing applied to the maps.
\label{tbl-2}}
\tablewidth{0pt}
\tablehead{
 \colhead{Band Center (GHz)} & \colhead{$-20^o < b < -10^o$ } & 
 \colhead{$-30^o < b < -20^o$ } & 
 \colhead{$-40^o < b < -30^o$} & \colhead{$-50^o < b < -40^o$}
& \colhead{$-60^o < b < -50^o$}    
}
\startdata
 410 & $(680 \pm 40)$  & $(280 \pm 90)$  &  $(180 \pm 60)$  &  $(110 \pm 35)$  &  $(120 \pm 40)$  \\
 240 & $(54 \pm 10)$   & $(15 \pm 3)$    &  $(9.6 \pm 2.0)$ &  $(6.0 \pm 1.2)$ &  $(6.4 \pm 1.3)$      \\
 150 & $(20 \pm 5)$    & $(2.8 \pm 1.7)$ &  $(1.7 \pm 1.1)$ &  $(1.1 \pm 0.7)$ &  $(1.2 \pm 0.7)$      \\
  90 & $(12 \pm 10)$   & $ < 12 $ $(2\sigma)$ & $ < 7 $ $(2\sigma)$ & $< 5$ $(2\sigma)$&  $<5$ $2\sigma)$  \\
\enddata
\end{deluxetable}

\end{document}